\documentclass[11pt]{article}
	
	\newcommand{\blind}{0}
	
	\usepackage[margin=1in]{geometry}
    \usepackage{titlesec}
    
    \titleformat{\section}{\Large\bfseries}{\thesection}{1em}{}
    \titleformat{\subsection}{\large\bfseries}{\thesubsection}{1em}{}
    \titleformat{\subsubsection}{\normalsize}{\thesubsubsection}{1em}{}
	\usepackage{amsmath}
    \usepackage{amssymb}
	\usepackage{graphicx}
	\usepackage{enumerate}
	\usepackage{xcolor}
    \usepackage{float}
    \usepackage{booktabs}
    \usepackage{siunitx}
    \usepackage{hyperref}
	\usepackage{natbib} %comment out if you do not have the package
	\usepackage{url} % not crucial - just used below for the URL
    \usepackage{bm}
\begin{document}
		
			%%%%%%%%%%%%%%%%%%%%%%%%%%%%%%%%%%%%%%%%%%%%%%%%%%%%%%%%%%%%%%%%%%%%%%%%%%%%%%
		\def\spacingset#1{\renewcommand{\baselinestretch}%
			{#1}\small\normalsize} \spacingset{1}
		%%%%%%%%%%%%%%%%%%%%%%%%%%%%%%%%%%%%%%%%%%%%%%%%%%%%%%%%%%%%%%%%%%%%%%%%%%%%%%
		
		\if0\blind
		{
			\title{\bf Estimating Decision Uncertainty from Preference Uncertainty: Application to Ground Vehicle Design}
			\author{Chia-Ruei Liu $^a$, Yongjia Song $^a$, Qiong Zhang $^b$, and Cameron Turner $^c$\\
			$^a$ Department of Industrial Engineering, Clemson University, Clemson, SC, USA \\
             $^b$ School of Mathematical and Statistical Sciences, Clemson University, Clemson, SC, USA \\
              $^c$ Department of Mechanical Engineering, Clemson University, Clemson, SC, USA}
			\date{}
			\maketitle
		} \fi
		
		\if1\blind
		{

            \title{\bf \emph{IISE Transactions} \LaTeX \ Template}
			\author{Author information is purposely removed for double-blind review}
			
\bigskip
			\bigskip
			\bigskip
			\begin{center}
				{\LARGE\bf \emph{IISE Transactions} \LaTeX \ Template}
			\end{center}
			\medskip
		} \fi
		\bigskip
		
	\begin{abstract}
    Engineering design problems are often modeled as multi-objective optimization tasks in which a scalarized utility function selects an optimal design from the Pareto set. In practice, preferences are imperfectly known, so uncertainty in the preference model leads to uncertainty in the resulting optimal design. This paper proposes a probabilistic framework that treats preference parameters as random variables and examines how preference uncertainty propagates to decision uncertainty. A random preference vector induces a probability distribution over optimal designs, allowing us to identify which regions of the Pareto front are most likely to be selected and to assess recommendation stability under preference variability. To explain the sources of this variability, we apply variance-based global sensitivity analysis to the induced optimal solutions, using Sobol' indices and Shapley values to quantify the contributions of individual design variables and their dependencies. We further summarize the overall dispersion of the optimal-design distribution using the Fréchet variance, which provides a scalar measure of decision stability under a given preference model. Two vehicle design case studies demonstrate how problem structure can lead to discrete versus continuous decision distributions and show how the proposed quantities support preference-aware design analysis.     \footnote{DISTRIBUTION STATEMENT A. Approved for public release; distribution is unlimited. OPSEC10263}
	\end{abstract}
			
	\noindent%

	%\newpage
	\spacingset{1.5} % DON'T change the spacing!

\section{Introduction} \label{s:intro}

\begin{figure}[H]
    \centering
    \includegraphics[scale=0.6]{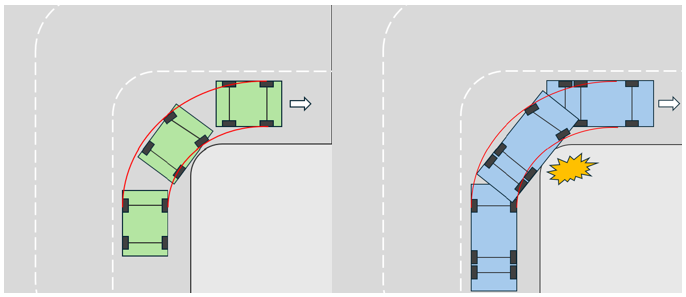}
    \caption{Geometric design choices in a ground vehicle and their effects on contact patch area and turning diameter. The left panel shows a more compact vehicle configuration, which is associated with a smaller contact patch area and a smaller turning diameter. The right panel shows a configuration with a larger contact patch area that requires a longer frame, resulting in a larger turning diameter. This illustrates the structural trade-off between contact patch area and turning diameter.}
\label{fig:vehicle_geometry}
\end{figure}

Engineering design problems are often formulated as multi-objective optimization problems \citep{hakanen2021multiobjective}. Designers must balance competing performance attributes such as cost, efficiency, safety, and reliability while satisfying engineering and regulatory constraints \citep{keeney1993decisions}. 
In the context of ground vehicle design \citep{de2022decomposition}, geometric choices in the configuration of the running gear and frame layout jointly influence key performance attributes such as back deck overhang, vehicle length, running gear contact patch area, and turning diameter. These objectives are structurally coupled and often move in opposite directions. For example, the running gear contact patch area  and the turning diameter are a pair of competing  attributes. The running gear contact patch area refers to the total area of contact between the running gear (e.g., wheels or tracks) and the ground, which is directly related to load distribution and terrain adaptability. Larger contact patch areas are generally preferred for stability and off-road performance. In contrast, the turning diameter characterizes the minimum circular path that the vehicle can follow during a turning maneuver, with smaller turning diameters indicating better maneuverability in confined environments. Increasing the contact patch area is typically associated with a longer running gear and frame, whereas reducing the turning diameter generally demands a more compact layout, as illustrated in Figure~\ref{fig:vehicle_geometry}. When there are competing objectives, multiple objective optimization methods typically can provide a set of solutions that are not worse than any other feasible alternatives, which is referred to as the Pareto set in the literature of multi-objective optimization \citep{deb2011multi}. Selecting a specific solution from the Pareto set typically depends on the preferences of the decision maker.
%Although the Pareto front provides a formal description of such trade-offs , it does not determine which design should be selected.

A decision maker's preference describes how they evaluate and trade off the competing objectives. A common approach is to represent preferences through scalarization or a utility function, such as a weighted-sum formulation, that assigns different weights 
to different objectives
%a single performance score to a vector of objective values 
\citep{keeney1993decisions, miettinen1999nonlinear}. Such utility-based models play a central role in Multiple Criteria Decision Analysis, where preferences determine how conflicts among objectives are resolved \citep{belton2012multiple}. With a fixed preference specification or weights for different objectives, classical multi-objective optimization reduces the problem to a scalar optimization task that selects the design maximizing the decision maker's utility. Therefore, the solutions provided in the Pareto set from multi-objective optimization can also be distinguished through the decision maker's utility scores. 
%However, although the Pareto front summarizes the set of non-dominated solutions, it does not determine which design should ultimately be chosen, as this decision depends entirely on the underlying preferences.

However, the main challenge is that preferences (or the weights of objectives) are rarely known precisely, as they are latent and cannot be directly observed. 
In practice, decision makers rarely provide explicit weights or utility functions with precision; their expressed preferences may vary with context, exhibit stochastic noise, and evolve as their understanding of trade-offs changes \citep{fisher2022bayesian, fisher2025approximate}.
%Decision makers often struggle to articulate explicit weights or utility functions, their priorities may shift, and their judgments exhibit noise and context dependence . 
As a result, many existing approaches either enumerate the entire Pareto set without using preferences \citep{deb2011multi, marler2004survey, de2022decomposition} or assume that a fixed and correctly specified preference model is available \citep{keeney1993decisions, ehrgott2005multicriteria}. Even preference-based optimization methods typically focus on inferring a single utility-maximizing design, rather than quantifying the variability in which designs may be selected under uncertain preferences \citep{astudillo2020multi}. These perspectives overlook an important source of uncertainty: uncertainty in preferences can propagate to uncertainty in the resulting optimal design even when the engineering model is deterministic. %From a practical standpoint, this means that relying on a single recommended design may be misleading if many alternative designs would be chosen under equally plausible preference specifications.

To fill this gap, we propose a framework that treats preferences as uncertain quantities and analyzes how preference uncertainty propagates to decision uncertainty. 
We model the preference parameters as random variables, which 
induces a probability distribution over the optimal solutions of the utility function. Thus, this framework transforms the classical Pareto analysis from a deterministic characterization into a probabilistic one. The resulting distribution provides actionable information about which regions of the Pareto set are likely to be selected under uncertain preferences and how sensitive these selections are to variations in preferences. 
Existing approaches related to preference elicitation and preference-guided optimization address uncertainty in different ways. D-optimal criteria aim to reduce uncertainty in the parameters of the preference model \citep{fisher2022bayesian, fisher2025approximate}, and acquisition strategies such as Expected Improvement under Utility Uncertainty identify designs with high expected utility under the current preference model \citep{astudillo2020multi}. By contrast, a probabilistic view of the optimal design reveals whether the current preference information is sufficient to support a stable design recommendation. A highly dispersed decision distribution signals that additional preference information may be needed, while a concentrated distribution suggests limited benefit from additional preference elicitation. Although elicitation is outside the scope of this study, the quantities developed here establish the foundations for such future extensions.

This paper contributes to the literature in four key aspects. First, we formulate a general optimization framework in which uncertain preferences induce a probability distribution over optimal design decisions, thereby extending classical multi-objective optimization from a deterministic to a probabilistic setting. Second, we show that the structure of the optimization problem shapes the form of decision uncertainty. Linear versus nonlinear trade-offs lead to qualitatively different preference-induced outcomes, ranging from discrete to continuous distributions over the Pareto set. Third, we integrate variance-based global sensitivity analysis, based on Sobol' indices~\citep{saltelli2008global} and Shapley values~\citep{owen2017shapley}, to quantify how design variables and their dependencies contribute to the variability in optimal designs, providing interpretable measures of design importance under preference uncertainty. Finally, we introduce the Fréchet variance~\citep{frechet1948elements, dubey2019frechet} as a scalar measure that summarizes the overall dispersion of the induced decision distribution. %While preference elicitation is beyond the scope of this study, this measure provides a natural basis for future elicitation strategies aimed at reducing decision uncertainty.

The remainder of the paper is organized as follows: In Section~\ref{s:literature}, we review related work on multi-objective optimization, Bayesian preference learning, preference-based optimization, and global sensitivity analysis. In Section~\ref{s:methods}, we present the proposed framework for propagating preference uncertainty to decision uncertainty and introduce the associated Sobol' and Shapley sensitivity measures together with the Fréchet variance as a summary indicator of decision dispersion. In Section~\ref{s:numerical}, we introduce vehicle design case studies and report numerical results on the induced decision distributions and their sensitivities. In Section~\ref{s:conclusion}, we conclude with a discussion of practical implications and directions for future research.

\section{Literature Review} \label{s:literature}

This section reviews the related literature on multi-objective optimization, Bayesian preference learning, preference-based optimization, and global sensitivity analysis.

Classical multi-objective optimization provides a mathematical framework for analyzing trade-offs among conflicting performance attributes by identifying the set of Pareto-efficient solutions \citep{marler2004survey, deb2011multi}. A wide range of methods has been developed to construct or approximate the Pareto set, including scalarization-based formulations \citep{miettinen1999nonlinear} and evolutionary algorithms designed to approximate diverse sets of non-dominated solutions \citep{deb2002fast}. These approaches are effective at identifying non-dominated solutions and characterizing the attainable trade-off structure, but they generally assume that user preferences are known a priori and fixed throughout the optimization process. As a result, they do not quantify how uncertainty or ambiguity in preference specification may influence which solutions are ultimately selected.  

Beyond classical multi-objective optimization, multi-objective Bayesian optimization (MOBO) provides a sample-efficient approach for exploring and modeling the Pareto set when function evaluations are costly or simulation-based \citep{knowles2006parego, feliot2017bayesian, garrido2019predictive}. Recent entropy- and uncertainty-based MOBO algorithms further improve sampling efficiency by prioritizing evaluations that reduce uncertainty around the Pareto set \citep{belakaria2019max, belakaria2020uncertainty}. These methods build surrogate models over multiple objectives and use acquisition functions to guide exploration toward non-dominated regions. Although MOBO improves efficiency and provides posterior uncertainty quantification over objective values, its primary goal remains the characterization or approximation of the Pareto set. Consequently, existing MOBO methods do not explicitly model or quantify how uncertainty in a decision maker's preferences propagates to uncertainty in the resulting design recommendations.

Preferential and utility-aware optimization methods attempt to align the search process with 
the preference of a decision maker by learning from pairwise or ranking-based feedback \citep{brochu2010tutorial, gonzalez2017preferential, astudillo2020multi, lin2022preference}. These approaches, which operate within a black-box optimization framework where the latent utility function must be inferred from indirect preference signals, reduce the reliance on a fully specified preference model and avoid exhaustive Pareto exploration, and are typically used to identify one or a small set of high-utility designs from which the decision maker makes a final choice. However, they do not provide a probabilistic characterization of how uncertainty in the inferred preference model propagates to uncertainty over which design would ultimately be selected, nor of how likely alternative designs are under different preference variations.

Although identifying how preference uncertainty affects the selected design is valuable, it is also important to understand which design variables contribute the most to such variability to support interpretable and trustworthy decision analysis. In this regard, global sensitivity analysis (GSA) offers a principled framework for attributing variability in a model output to uncertainty in its inputs, with widely used approaches including variance-based Sobol' indices \citep{saltelli2008global} and Shapley-value formulations that remain valid under dependent inputs \citep{song2016shapley, owen2017shapley}. However, existing GSA applications primarily focus on analyzing uncertainty in physical or simulation-based performance outputs rather than uncertainty in the optimal decisions themselves \citep{iooss2015review, razavi2021future}. To the best of our knowledge, the use of GSA as a diagnostic tool for explaining preference-induced decision variability has not been explicitly examined.

% Finally, statistical summaries based on the Fréchet mean and Fréchet variance provide natural extensions of classical moment-based quantities to general metric spaces. The Fréchet mean was originally introduced as the minimizer of the expected squared distance in an arbitrary metric space \citep{frechet1948elements}, and its asymptotic properties have since been studied in manifold-valued settings \citep{bhattacharya2003large}. More recently, Fréchet-based measures have been used to extend analysis-of-variance ideas to non-Euclidean data objects \citep{dubey2019frechet}. Their generality makes them well suited for applications where the underlying decision space may not be intrinsically Euclidean, although in this work the design variables lie in a Euclidean space. Here, the Fréchet variance is used as a scalar indicator of the overall dispersion of the optimal design distribution, complementing the sensitivity analyses described above.

\section{Methodology} \label{s:methods}
We start with introducing a general model and notation for propagating preference uncertainty through a multi-objective optimization problem. Consider an input vector $\bm x \in \mathcal{X} \subseteq \mathbb{R}^d$ and $m$ objective functions collected as $\bm f(\bm x) =  (f_1(\bm x),\dots,f_m(\bm x))$. A preference vector $\bm \beta \in \mathbb{R}^m$ specifies how the decision maker weights the objectives, and the corresponding scalarized utility \citep{keeney1993decisions, miettinen1999nonlinear} is defined as \[ U_{\bm \beta}(\bm x) = \bm \beta^\top \bm f(\bm x) \] For a fixed preference vector $\bm\beta$, the best design is the solution of the scalar optimization problem \[ \bm x^*(\bm \beta) \in \arg\min_{\bm x \in \mathcal{X}}U_{\bm \beta}(\bm x) \]

\begin{figure}[H]
    \centering
    \includegraphics[width=1\textwidth]{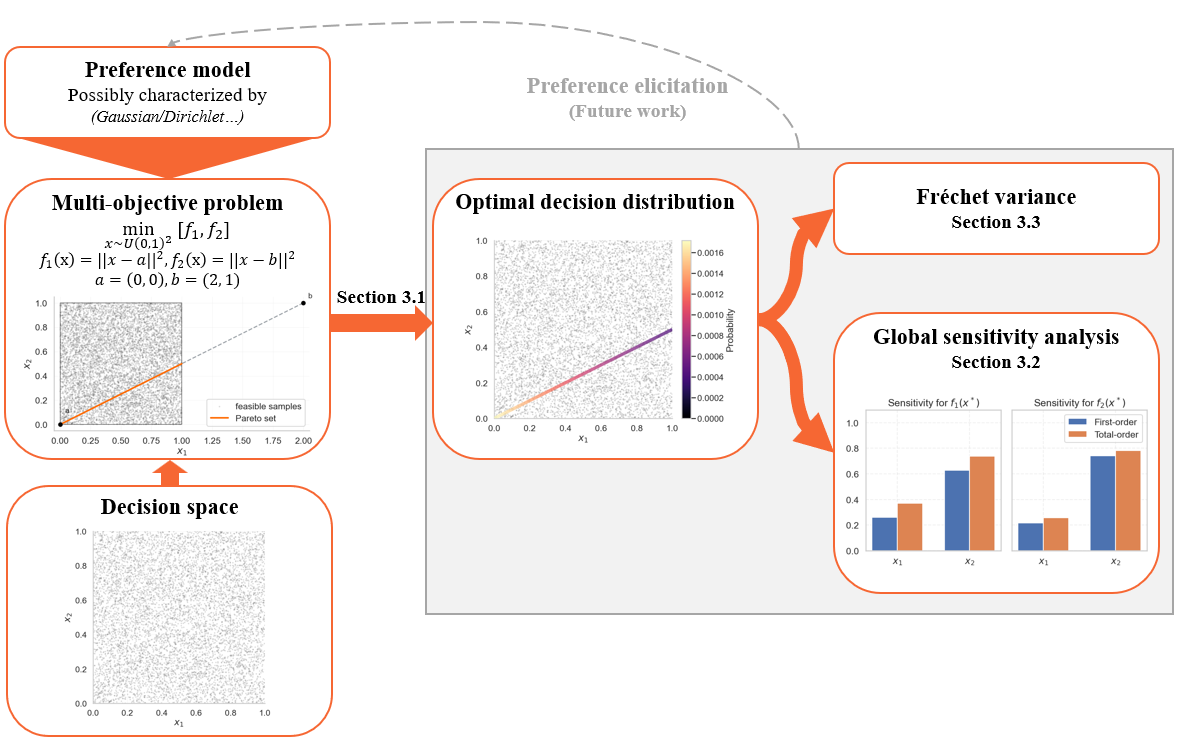}
    \caption{An overview of the proposed framework
 on propagating preference uncertainty to decision uncertainty. }
\label{fig:pipeline}
\end{figure}

In practice, preferences are rarely known precisely. We model $\bm \beta$ as a random vector with the probability density function $p(\bm \beta)$. This induces a random optimal decision $\bm X = \bm x^*(\bm \beta)$ and therefore a probability distribution over the input space $\mathcal X$. Our aim is to characterize this induced distribution and to understand how uncertainty in preferences propagates to uncertainty in the optimal decision. 
An overview of the proposed framework is
shown in Figure~\ref{fig:pipeline}.

As a simple geometric illustration, consider a bi-objective problem where the input vector is $\bm x=(x_1,x_2)\in\mathbb{R}^2$. The objectives are:
\[f_1(\bm x)=\|\bm x- \bm a\|^2\quad f_2(\bm x)=\|\bm x-\bm b\|^2 
\quad\mathrm{with}\quad \bm a=(0,0)\quad\mathrm{and} \quad\bm b=(2,1).
\]
As shown in Figure \ref{fig:pipeline}, 
the Pareto-optimal solutions for this problem lie along the straight line segment connecting $\bm a$ and $\bm b$, and different values of the preference vectors $\bm \beta$ select different points on this segment. As a result, uncertainty of the preference $\bm \beta$ induces a probability distribution of the optimal decision over this segment, offering a clear geometric illustration of how preference uncertainty transforms a deterministic Pareto set into a probabilistic one. 

Although the induced distribution of $\bm X$ shows which regions of the Pareto-optimal set are most likely under uncertain preferences, it does not explain why the recommended designs vary across preference realizations. Therefore, we use global sensitivity analysis to attribute the variability in the induced optimal inputs and their objective outcomes to individual input variables and their dependencies. This can identify the main drivers of decision uncertainty. These attributions provide interpretable guidance on where input changes affect the results the most. To complement these variable-level insights, we also report the Fréchet variance as a single scalar summary of the overall dispersion of the induced optimal-solution distribution. 

\subsection{Propagating preference uncertainty to decision uncertainty} \label{s:methods.1}
Uncertainty in preferences propagates through optimization and results in uncertainty in the solutions. When preferences are modeled as random variables, each realization of the preference vector $\bm \beta$ defines a set of minimizers $\mathcal{X}^*(\bm \beta) = \arg\min_{\bm x \in \mathcal{X}} U_{\bm \beta}(\bm x)$. In principle, multiple optimal solutions may exist for the same $\bm\beta$, particularly in linear programs or in nonlinear problems with flat regions (see Supplementary Material Section~S.1). In practice, our implementation records the single optimal solution returned by the solver~\citep{gurobi} for each sampled preference realization. Accordingly, the mapping $\bm\beta \mapsto \bm x^*(\bm\beta)$ should be interpreted as selecting one representative optimizer rather than explicitly enumerating all possible ties. As a result, a distribution over preferences induces a corresponding distribution over optimal design solutions. We next illustrate this propagation under two representative cases of preference uncertainty. For clarity of exposition, the illustrative examples in the following subsections are presented in maximization form. These formulations are equivalent to the minimization framework defined above.

\subsubsection{Discrete solution outcomes} \label{s:methods.1.1}
%For notational convenience, the following illustrative examples are presented in maximization form, without loss of generality. 
We first illustrate a preference-to-decision uncertainty propagation example when the decision uncertainty is represented as a discrete probability distribution. Consider a single decision variable $x \in [0,1]$ with two linear objective functions: 
\[ f_1(x) = a_1 + b_1 x, \quad f_2(x) = a_2 + b_2 x. \]
Given a preference vector $\bm \beta=(\beta_1,\beta_2)^\top \sim \mathcal{N}(\bm \theta,\bm \Sigma)$, 
the corresponding utility can be written as
\[ U_{\bm \beta}(x) = \beta_1 f_1(x) + \beta_2 f_2(x) = c + (\beta^\top \bm b)x, \quad \bm b = \begin{pmatrix} b_1 \\ b_2 \end{pmatrix}. \]
Maximizing the utility over $x \in [0,1]$ yields the optimal decision
\[
x^*(\bm \beta) =
\begin{cases}
1, & \beta^\top \bm b \geq 0, \\[0.3em]
0, & \beta^\top \bm b < 0,
\end{cases}
\]
which shows that the solution space collapses to two discrete outcomes, $\{0,1\}$. 
The probability of selecting $x^*=1$ is then given by
\[
\mathbb{P}(x^*=1) = 
\Phi\!\left( \frac{\bm b^\top \bm \theta}{\sqrt{\bm b^\top \bm \Sigma \bm b}} \right),
\]
where $\Phi(\cdot)$ denotes the standard normal cumulative distribution function. 
This closed-form expression explicitly demonstrates how uncertainty in $\beta$ translates into a probability distribution over decisions. 

\begin{figure}[h]
\centering
\includegraphics[width=1\textwidth]{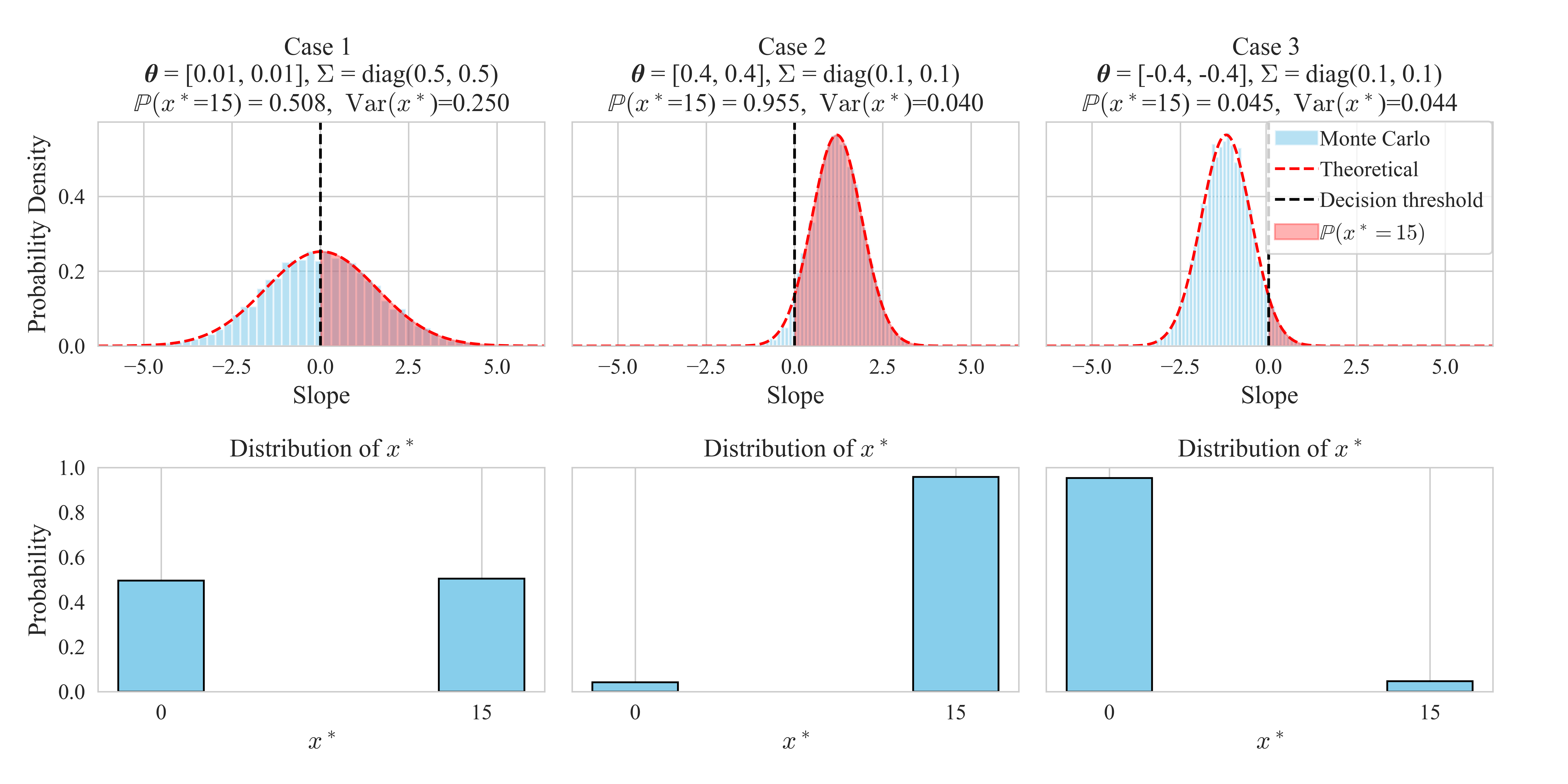}
\caption{Discrete example: distributions of optimal decisions under three preference scenarios.}
\label{fig:pref_ex1}
\end{figure}

To further illustrate the same mechanism with different numerical scales, we specify the coefficients in the linear objectives as \[ \bm a = \begin{pmatrix} a_1 \\ a_2 \end{pmatrix} =  \begin{pmatrix} -15 \\ -55 \end{pmatrix} , \quad \bm b = \begin{pmatrix} b_1 \\ b_2 \end{pmatrix} = \begin{pmatrix} 1 \\ 2 \end{pmatrix},  \] with the decision variable defined over $x\in[0,15]$. 
% To further illustrate the same mechanism with different numerical scales, consider a simple bi-objective optimization problem
% \[
% \max_{x \in [0,15]} \ [f_1(x), \ f_2(x)], 
% \quad \text{where} \quad f_1(x) = -15 + x, \quad f_2(x) = -55 + 2x.
% \]
Under $\bm \beta \sim \mathcal{N}(\bm \theta,\bm \Sigma)$, we evaluate several preference scenarios. 
As shown in Figure~\ref{fig:pref_ex1}, different mean and covariance structures of $\beta$ lead to distinct distributions over the optimal decisions $x^*(\bm \beta)$, 
demonstrating the role of preference uncertainty in shaping decision outcomes. The top row shows the distribution of the term $\bm \beta^\top \bm b$, which  determines whether the optimal decision is $x^*=0$ (negative slope) or $x^*=15$ (positive slope). The blue histogram represents the empirical distribution obtained via Monte Carlo sampling of $\bm \beta$, while the red dashed curve denotes the corresponding theoretical density. The vertical dashed line marks the zero threshold at which the preferred decision switches. The bottom row reports the resulting probability mass of the optimal solution, $\mathbb{P}(x^*=0)$ and $\mathbb{P}(x^*=15)$, obtained from the same Monte Carlo samples. Across the three cases, both Case~1 and Case~2 tend to select $x^*=15$, but Case~1 exhibits higher variability because its preference distribution has larger covariance. In contrast, Case~2, with smaller covariance, yields a tightly concentrated distribution and becomes nearly deterministic. Case~3 mirrors Case~2 with the opposite mean direction, leading to a similarly concentrated choice of $x^*=0$. These results demonstrate that the variability of the optimal decision is directly influenced by the level of uncertainty in the preference distribution. 

\subsubsection{Continuous solution outcomes} \label{s:methods.1.2}
We next consider when the decision uncertainty is represented as a continuous probability distribution. Let $x \in \mathbb{R}$ be a single decision variable and consider two concave quadratic objective functions:
\[
f_1(x) = a_1 - b_1 x - x^2, 
\qquad 
f_2(x) = a_2 - b_2 x - x^2.
\]
For a preference vector $\bm \beta = (\beta_1, \beta_2)^\top$, the utility is
\[
U_{\bm \beta}(x) = \beta_1 f_1(x) + \beta_2 f_2(x) 
= c - (\bm \beta^\top \bm b) x - (\beta_1 + \beta_2) x^2,
\quad 
\bm b = \begin{pmatrix} b_1 \\ b_2 \end{pmatrix}.
\]
Throughout this subsection, we assume that the preference weights satisfy $\beta_1 + \beta_2 > 0$ almost surely. Maximizing the utility yields a unique optimum with a closed-form solution:
\[
x^*(\bm \beta) = -\frac{\bm \beta^\top \bm b}{2(\beta_1+\beta_2)}.
\]
Cases with $\beta_1 + \beta_2 \le 0$ are excluded under this modeling assumption.

\begin{figure}[h]
\centering
\includegraphics[width=1\textwidth]{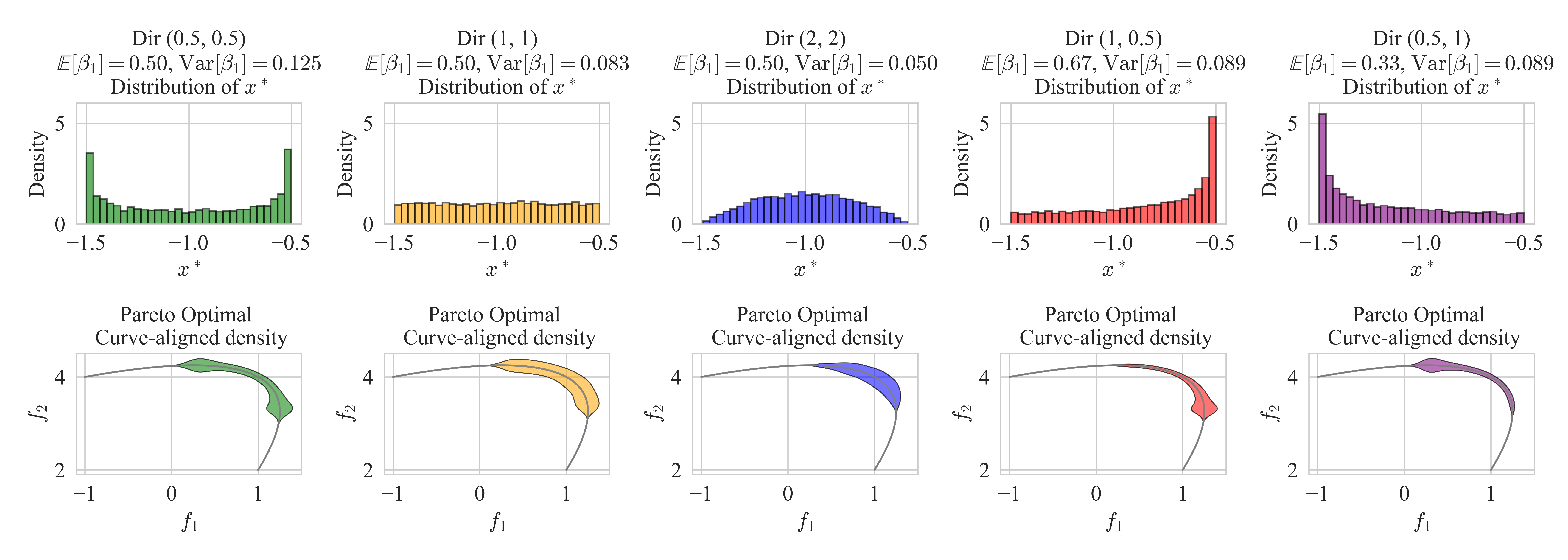}
\caption{Continuous example: distributions of optimal decisions under five preference scenarios.}
\label{fig:pref_ex2}
\end{figure}

In contrast to the discrete outcome case, here the optimal decision varies continuously with $\bm \beta$, resulting in a continuous distribution of $x^*(\bm \beta)$. To illustrate this continuous behavior, consider the following example:
\[
\max_{x \in \mathbb{R}} \ [f_1(x), \ f_2(x)], 
\quad \mathrm{where} \quad
f_1(x) = 1 - x - x^2, 
\quad 
f_2(x) = 2 - 3x - x^2.
\]
Assume that $\bm \beta$ follows the Dirichlet distribution
$\mathrm{Dir}(\alpha_1,\alpha_2)$. We evaluate five preference scenarios. 
Figure~\ref{fig:pref_ex2} shows how different Dirichlet parameters produce distinct continuous distributions of optimal decisions, illustrating how preference uncertainty propagates to decision uncertainty. The top row shows the empirical distribution of $x^*(\bm \beta)$ obtained from Monte Carlo sampling of $\bm \beta$ under five different Dirichlet preference settings. The bottom row presents the corresponding curve–aligned density along the Pareto front $(f_1(x), f_2(x))$, indicating how the probability mass over $x^*(\bm \beta)$ maps into trade-off space. Across the five scenarios, the first three cases share the same mean preference direction $(\mathbb{E}[\beta_1]=\mathbb{E}[\beta_2]=0.5)$ but differ in concentration levels. As the Dirichlet parameters increase from $\mathrm{Dir}(0.5,0.5)$ to $\mathrm{Dir}(2,2)$, preference uncertainty decreases, leading to progressively more concentrated decision distributions $x^*(\bm \beta)$ and reduced dispersion of probability mass along the Pareto front. In contrast, the other two settings $(\mathrm{Dir}(1,0.5)$ and $\mathrm{Dir}(0.5,1))$ shift the mean preference direction toward opposite objectives, which results in decision distributions that favor different regions of the feasible domain and align with different segments of the Pareto-optimal trade-off curve. These results show that preference concentration governs decision variability, while preference mean dictates where probability mass accumulates along the Pareto front. 

The preceding lower dimensional illustrations help clarify the mechanisms by which preference uncertainty induces uncertainty in optimal decisions. In real engineering applications, however, the decision space is high dimensional, and identifying which variables and interactions drive this uncertainty becomes substantially more challenging. To address this interpretability challenge, we next introduce a global sensitivity analysis framework.

\subsection{Global sensitivity analysis} \label{s:methods.2}
To analyze how preference uncertainty propagates to variability in the selected designs, we evaluate the sensitivity of each objective outcome to variations in the optimal solution. Let $Y_\ell = f_\ell(\bm X), \quad \ell = 1,\dots,m, $ denote the objective values induced by the random optimal solution $\bm X = \bm x^*(\bm \beta)$. Global sensitivity analysis (GSA) provides a principled way to attribute the variability of each $Y_\ell$ to the individual components of $\bm X$ and their interactions, thereby identifying which design variables drive decision uncertainty. 
Since the components of the optimal solution $\bm X$ are generally dependent as a result of the optimization mapping from preferences to decisions, variance attribution methods that assume independent inputs are no longer valid. Accordingly, we employ a variance-based sensitivity analysis framework, using Sobol' indices for comparison in the independent-input setting \citep{saltelli2008global} and Shapley values as the primary measure when dependencies are present \citep{owen2017shapley}. Detailed formulations and estimation procedures are provided in the Supplementary Material (Section~S.2). We next introduce a scalar measure that summarizes the overall dispersion of the induced optimal-solution distribution.

\subsection{Decision uncertainty measure}\label{s:methods.3}

As a scalar summary of the overall dispersion of the optimal solution distribution, we adopt the Fréchet variance \citep{frechet1948elements, bhattacharya2003large}, defined for a general metric space $(\mathcal{X}, d)$ as
\[
\mathrm{Var}_F(\bm X) = \min_{\bm z \in \mathcal{X}} 
\mathbb{E}\big[d(\bm X, \bm z)^2\big],
\]
where $\bm X$ is the random optimal decision, and the expectation
is taken with respect to the distribution of $\bm X$. 
The minimizer is the Fréchet mean. Since our decision space is Euclidean and we use the standard $\ell_2$ distance, $d(\bm x, \bm z)=\|\bm x- \bm z\|_2$, the Fréchet mean coincides with the arithmetic mean of the optimal solutions, and the Fréchet variance reduces to the usual variance under this metric. All computations are performed in the normalized decision space so that each design variable contributes comparably to the distance. A small Fréchet variance indicates that most preference realizations lead to similar designs, whereas a larger value reflects substantial decision variability. This measure is reported in the case study as an overall indicator of design stability under preference uncertainty.

\section{Case study} \label{s:numerical}
In this section, we assess the performance of the proposed approach using two variants of the same vehicle design problem, one leading to a discrete probability distribution over a finite set of design decisions and the other leading to a continuous probability distribution. To provide a clear basis for evaluation, we first describe the formulation of these problems, including modeling assumptions, decision variables, objective functions, and constraints. We then present how the original problems can be reformulated, for example, through epigraph representations and normalization, in order to make them compatible with the proposed optimization framework.

\subsection{Vehicle design problem}
The case study is based on a vehicle design task that involves six decision variables:
\begin{itemize}
    \item $x_1$: Running gear back mount inset (in), $x_1 \in [4.5,\ 23]$
    \item $x_2$: Running gear front mount inset (in), $x_2 \in [6,\ 20]$
    \item $x_3$: Running gear road wheel radius (in), $x_3 \in [1,\ 5]$
    \item $x_4$: Running gear drive gear radius (in), $x_4 \in [1,\ 3]$
    \item $x_5$: Frame axle mount inset (in), $x_5 \in [0,\ 15]$
    \item $x_6$: Frame underbody length (in), $x_6 \in [34,\ 132]$
\end{itemize}

Each candidate design is evaluated according to four objectives:
\begin{itemize}
    \item $f_1$: Back deck overhang (inches),
    \item $f_2$: Vehicle length (inches),
    \item $f_3$: Running gear contact patch area (in$^2$, depends on gear type),
    \item $f_4$: Curb-to-curb turning diameter (inches, depends on the steering type).
\end{itemize}

To reduce unnecessary complexity while preserving the essential design trade-offs, the following assumptions are imposed:
\begin{itemize}
    \item The gear type is fixed to a trapezoidal track configuration. This ensures a consistent geometry and avoids additional variables associated with alternative track layouts.
    \item The constraint $x_4 \geq x_3$ (drive gear radius $\ge$ road wheel radius) is imposed for mechanical feasibility and to eliminate case distinctions in $f_3$ and $f_4$.
    \item The track width is fixed at $14$ inches. This prevents additional nonlinearities in $f_3$ and $f_4$.
    \item Steering is restricted to either Ackermann or pivot/skid configurations. These two cases yield fundamentally different solution structures: a discrete probability distribution over potential optimal decisions for Ackermann and a continuous probability distribution over potential optimal decisions for pivot/skid.
\end{itemize}

The optimization goal is:
\[ \min \; [\, f_1, \; f_2, \; -f_3, \; f_4 \,]. \]

\subsection{Problem formulation}

Both case studies are based on the same underlying vehicle design optimization problem. 
The decision vector includes six continuous variables $(x_1,\dots,x_6)$ and two auxiliary epigraph variables $(\theta_1,\theta_2)$. 
The objective is a preference-weighted utility function
\[
\min_{x_1, x_2, x_3, x_4, x_5, x_6,\;\theta_1,\theta_2} \; 
U(\bm x,\bm \theta, \bm \beta) 
= \beta_1 \tilde f_1 + \beta_2 \tilde f_2 - \beta_3 \tilde f_3 + \beta_4 \tilde f_4, 
\qquad \bm \beta \sim \mathcal{TMVN}(\bm \mu, \bm \Sigma),
\]
where $\mathcal{TMVN}(\bm \mu,\bm \Sigma)$ denotes a truncated multivariate normal distribution~\citep{botev2017normal} with mean $\bm \mu$ and covariance $\bm \Sigma$, truncated such that $\beta_i \geq \epsilon$ for all $i$, with $\epsilon > 0$ a small constant that rules out nonpositive weights, and the normalized objectives $\tilde f_i$ are given by:
\[
\tilde f_1=\frac{\theta_1}{23}, 
\quad \tilde f_2=\frac{x_6 + \theta_1 + \theta_2}{141}, 
\quad \tilde f_3=\frac{\big(x_6-2x_5-3.414\,x_3-0.586\,x_4\big)14}{1792}.
\]

The only difference between the two case studies lies in the fourth objective $\tilde f_4$:  
\begin{equation}\label{eq:f4}
\tilde f_4 =
\begin{cases}
\dfrac{x_6-2x_5 - 4}{128}, & \text{Case Study 1 (Ackermann steering type)} \\
\dfrac{\big[(x_6-2x_5-3.414\,x_3-0.586\,x_4)14\big]^2-1936}{3211264}, & \text{Case Study 2 (Pivot/Skid steering type)}.
\end{cases}
\end{equation}

We normalize all objective functions to ensure that the preference weights $\bm \beta$ reflect relative trade-offs rather than being dominated by differences in numerical scale. Without normalization, objectives with larger magnitudes could disproportionately influence the scalarized utility, leading to a distorted interpretation of $\bm \beta$ and biased decision outcomes. For each objective, the scaling parameters are determined based on the minimum and maximum values attained over the same feasible region. This normalization plays the same role as a standard min--max transformation,
\[
\tilde f_i = \frac{f_i - f_i^{\min}}{f_i^{\max} - f_i^{\min}},
\]
which places all objectives on a comparable unit range while preserving their relative variation across the feasible space.

The constraints are identical across both cases:
\[
x_4 \geq x_3, \quad 
\theta_1 \geq x_1 - x_5, \quad 
\theta_2 \geq x_2 - x_5, 
\]
with variable bounds
\[
\theta_1 \in [0,23], \; \theta_2 \in [0,20], \
x_1 \in [4.5,23], \; x_2 \in [6,20], \; x_3 \in [1,5], \;
x_4 \in [1,3], \; x_5 \in [0,15], \; x_6 \in [34,132],
\]
and the feasibility condition
\[
x_6 - 2x_5 - 3.414\,x_3 - 0.586\,x_4 \geq 0 
\quad \text{(Running gear contact patch area $\geq 0$)}.
\]

In practical terms, Case Study 1 corresponds to trapezoidal-track vehicles equipped with Ackermann steering, for which the curb-to-curb turning diameter can be approximated by a linear function of the design variables. Since all objectives in Case Study 1 are affine, the scalarized problem is a linear program. For linear programs, whenever the optimal value is finite, at least one optimal solution occurs at an extreme point of the feasible polyhedron. Since the feasible polyhedron has only finitely many extreme points, all candidate optimal designs under preference uncertainty can be restricted to this finite set. By contrast, Case Study 2 corresponds to vehicles with pivot/skid steering, as found in tracked platforms. In this case, the curb-to-curb turning diameter is modeled by a quadratic expression in the design variables, reflecting the nonlinear geometry of skid-based turning. Since $\tilde f_1$, $\tilde f_2$, and $\tilde f_3$ are affine, $\tilde f_4$ is convex quadratic, and all constraints are linear, each scalarized problem $\min_x \sum_{i=1}^4 \beta_i \tilde f_i(x)$ is a convex quadratic problem. We solve these convex quadratic problems using Gurobi, which guarantees global optimality up to numerical tolerances \citep{gurobi}. Unlike the linear programming structure in Case Study 1, the optimal solutions in Case Study 2 form a continuous distribution rather than a finite set of extreme points.

\subsection{Experimental setup}
In both cases we draw $N=1000$ preference vectors $\bm \beta \sim \mathcal{TMVN}(\bm \mu, \bm \Sigma)$, where $\bm \mu=(1,1,1,1)$ and $\bm \Sigma=0.5\,\mathbf{I}_4$. We adopt a truncated multivariate normal distribution to enforce strictly positive preference weights, consistent with the interpretation of $\bm \beta$ as attribute importance weights in the vehicle design context. Allowing nonpositive weights would contradict the interpretation of $\bm \beta$ as preference intensities and be inconsistent with the goals of the vehicle design problem. Each sampled preference vector is then solved via the corresponding reformulated optimization model to obtain a Pareto-optimal decision under stochastic preferences. The next subsections introduce the case-specific definitions of $\tilde f_4$ that give rise to the two contrasting case studies.

\subsection{Case Study 1: Discrete optimal decision distribution}
In Case Study 1, the fourth objective function in \eqref{eq:f4} is defined as
\[
\tilde f_4 = \frac{x_6 - 2x_5 - 4}{128},
\]
which corresponds to the Ackermann steering configuration. This leads to an LP formulation of the overall problem. Since LP problems attain their optima at extreme points of the feasible polytope, the set of Pareto-efficient extreme design points is finite. Consequently, under preference uncertainty, the induced optimal decision distribution  places probability mass only on this finite collection of solutions, resulting in a discrete distribution (see Table~\ref{tab:lp_solution_distribution_sorted}).

Since the scalarized problems in this case are linear programs, a fixed preference vector $\bm\beta$ may still admit multiple optimal solutions due to structural degeneracy, for example when the optimum lies on a higher-dimensional face of the feasible polytope. Such multiplicities arise from the problem geometry rather than from the preference vector and therefore do not correspond to distinct preference-induced choices. In contrast to the preference-induced multiplicity discussed in the Supplementary Material (Section S.1), we use the solver’s returned extreme point as the representative design for each sampled $\bm\beta$, ensuring that the induced decision distribution and subsequent sensitivity and Fréchet analyses reflect preference uncertainty rather than artifacts of degeneracy. 

\begin{table}[H]
\centering
\caption{Discrete optimal decisions and their empirical probabilities from the LP under preference uncertainty ($N = 1000$)}
\label{tab:lp_solution_distribution_sorted}
\begin{tabular}{S[table-format=2.1]
                S[table-format=2.1]
                S[table-format=1.1]
                S[table-format=1.1]
                S[table-format=2.1]
                S[table-format=3.1]
                S[table-format=1.3]}
\toprule
{$x_1$} & {$x_2$} & {$x_3$} & {$x_4$} & {$x_5$} & {$x_6$} & {Probability} \\
\midrule
4.5 & 6.0 & 1.0 & 1.0 & 15.0 &  34.0 & \textbf{0.505} \\
4.5 & 6.0 & 1.0 & 1.0 &  6.0 &  34.0 & 0.186 \\
4.5 & 6.0 & 1.0 & 1.0 &  4.5 & 132.0 & 0.161 \\
4.5 & 6.0 & 1.0 & 1.0 &  4.5 &  34.0 & 0.121 \\
4.5 & 6.0 & 1.0 & 1.0 &  0.0 & 132.0 & 0.027 \\
\bottomrule
\end{tabular}
\end{table}

\begin{figure}[H]
    \centering
    \includegraphics[width=1\textwidth]{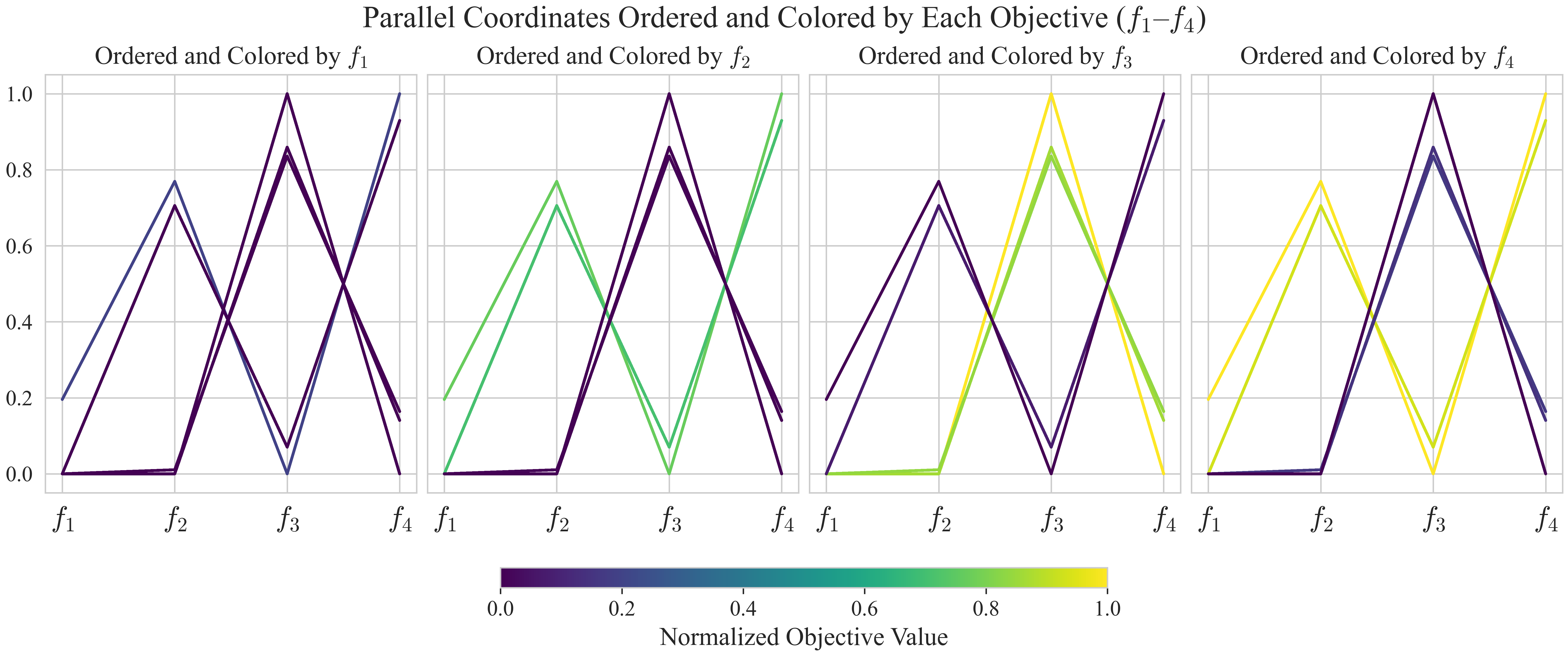}
    \caption{Parallel coordinates plot of normalized objective values for Pareto-optimal solutions in Case Study 1 (discrete distribution).}
\label{fig:preference_problem4_norm_parallel_coordinates}
\end{figure}

As shown in Figure~\ref{fig:preference_problem4_norm_parallel_coordinates}, the Pareto-optimal solutions exhibit clear trade-offs among the objectives. The effect of $f_1$ is relatively small, whereas $f_2$ and $f_4$ show an opposite trend with respect to $f_3$. Moreover, $f_2$ and $f_4$ follow a similar pattern, indicating consistent trade-off behavior.

\begin{figure}[H]
    \centering
    \includegraphics[width=1\textwidth]{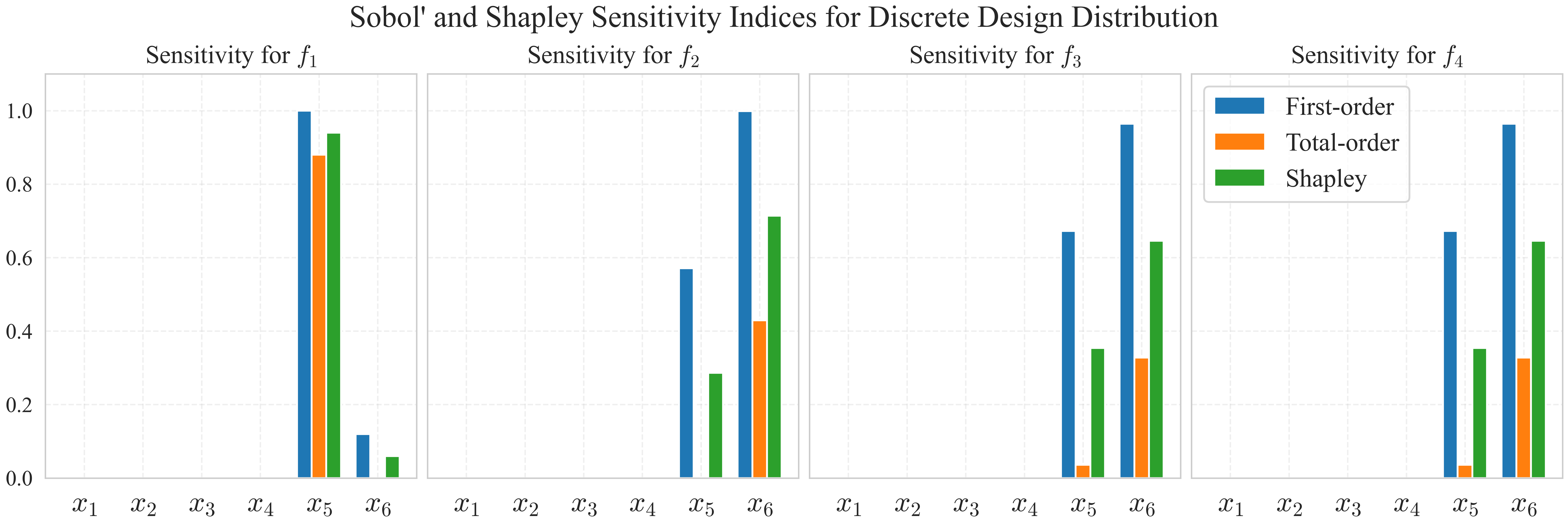}
    \caption{Sobol' and Shapley sensitivity indices for discrete optimal decision solutions across the four objectives.}
    \label{fig:preference_problem4_normalized_lp_sobol}
\end{figure}

\begin{figure}[H]
    \centering
    \includegraphics[width=1\textwidth]{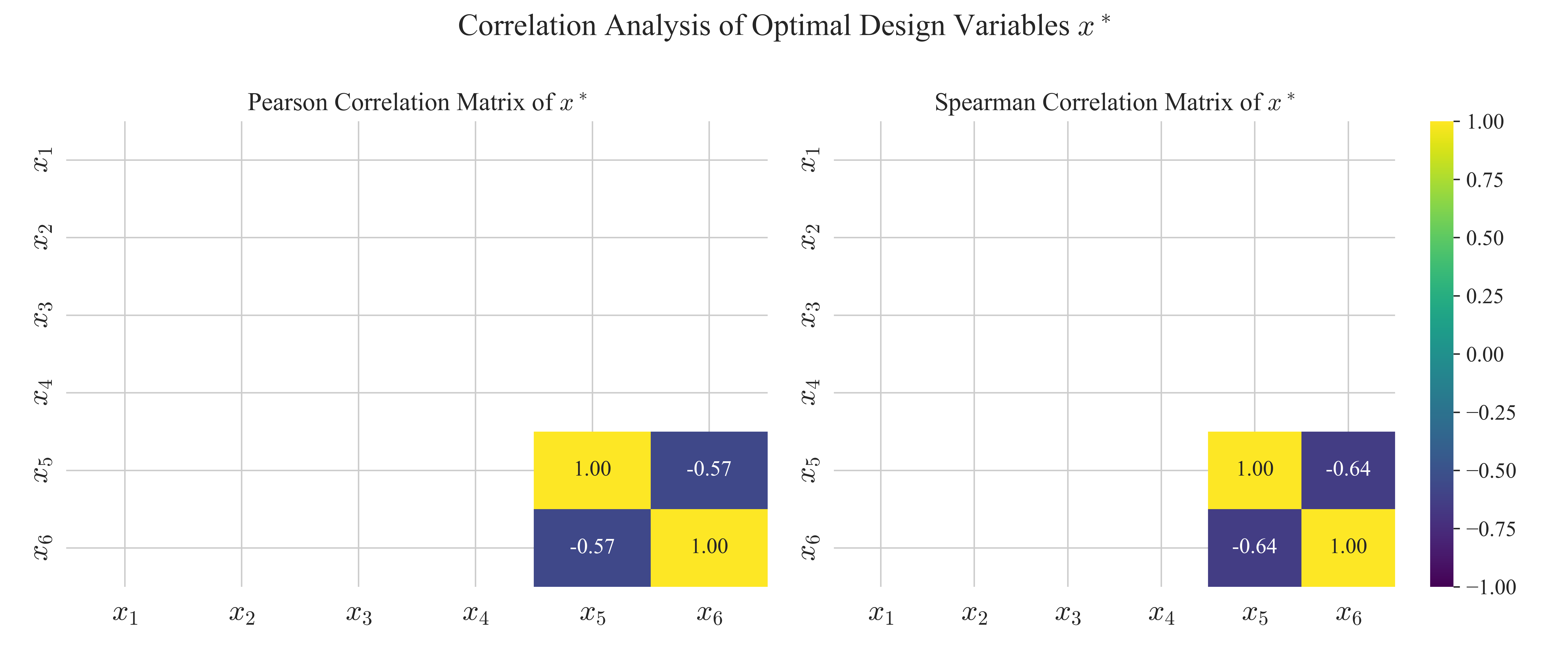}
    \caption{Pearson and Spearman correlation matrices of the discrete design distribution.}
    \label{fig:preference_problem4_norm_t_star_dependence}
\end{figure}

Figure~\ref{fig:preference_problem4_normalized_lp_sobol} reports the Sobol' and Shapley sensitivity indices of each objective value evaluated at the induced optimal designs. The results indicate that $x_5$ dominates all objectives, while $x_6$ has secondary influence and the other variables do not contribute. For $f_1$, $x_5$ is clearly the main factor, whereas $f_2$–$f_4$ are primarily driven by $x_5$ with some contribution from $x_6$. An anomaly arises when first-order Sobol' indices exceed the corresponding total-order values. Under independent inputs total-order indices should exceed or equal to their first-order counterparts, so this reversal points to dependencies among the design variables. Figure~\ref{fig:preference_problem4_norm_t_star_dependence} shows Pearson and Spearman correlation matrices, which reveal a moderate negative dependence between $x_5$ and $x_6$. This dependence may partly explain why some first-order Sobol' indices exceed their total-order counterparts. As Pearson and Spearman capture only linear or monotonic associations, more complex dependencies cannot be ruled out. In this setting, Shapley indices provide a more robust attribution of variable importance under correlated inputs.

\begin{figure}[H]
    \centering
    \includegraphics[width=1\textwidth]{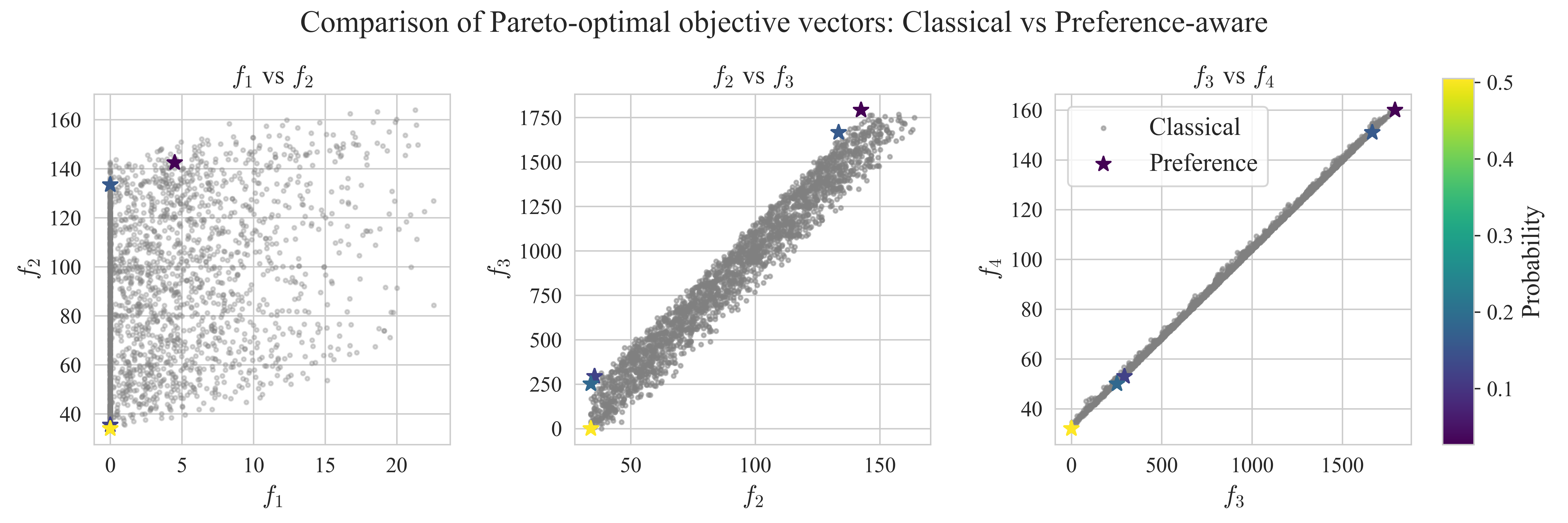}
    \caption{Classical Pareto-optimal objective vectors (grey) vs. preference-aware solutions (colored by probability mass) for discrete optimal decision.}
    \label{fig:preference_problem4_comparison_moo_preference}
\end{figure}

Figure~\ref{fig:preference_problem4_comparison_moo_preference} compares the results from classical and preference-aware optimization. For the classical baseline, we approximate the Pareto set by Latin Hypercube Sampling (LHS) of the design space followed by non-dominated sorting. Classical multi-objective optimization identifies the entire Pareto-optimal set, but it does not incorporate preferences and thus provides no indication of which solutions are more likely to be chosen under uncertainty. By contrast, our approach characterizes the induced decision distribution directly. In this LP case, the outcome is a discrete distribution over extreme-point Pareto-optimal designs. As illustrated in the figure, the probability mass is strongly concentrated on a small subset of extreme-point Pareto-optimal designs, while most other Pareto-efficient solutions have negligible probability. This pattern is consistent with the empirical distribution reported in Table~\ref{tab:lp_solution_distribution_sorted}, where a single design accounts for more than half of the total probability mass. Such concentration reveals that not all Pareto-efficient designs are equally likely to be selected under preference uncertainty, allowing decision makers to distinguish designs that are robustly favored from those that are sensitive to preference variation. 

While the probability mass over discrete optimal designs (Table~\ref{tab:lp_solution_distribution_sorted}) already reveals which designs are most likely to be selected under a given preference distribution, decision makers may also benefit from a single scalar measure that summarizes the overall stability of these recommendations. Importantly, the Fréchet variance is defined relative to a specific preference distribution, and different choices of $(\mu,\Sigma)$ lead to different levels of decision dispersion. To quantify how dispersed the induced decision distribution is, we report the Fréchet variance under several representative preference settings. A small Fréchet variance indicates that most preference realizations lead to very similar designs, whereas a large variance indicates that the resulting optimal designs are much more dispersed. Table~\ref{tab:frechet_variance_discrete} reports the values obtained under the different preference settings.

\begin{table}[H]
\centering
\caption{Fréchet variance of the discrete optimal decision distribution under different preference settings.}
\label{tab:frechet_variance_discrete}
\begin{tabular}{lccS[table-format=1.3]}
\toprule
Scenario & $\bm \mu$ & $\bm \Sigma$ & {Fréchet variance} \\
\midrule
Baseline          & $(1,1,1,1)$   & $0.5 I_4$  & 0.2719 \\
Low uncertainty   & $(1,1,1,1)$   & $0.1 I_4$  & 0.1302 \\
High uncertainty  & $(1,1,1,1)$   & $1.0 I_4$  & 0.3177 \\
Objective 3 emphasized & $(1,1,3,1)$   & $0.5 I_4$  & 0.1782 \\
Objective 4 emphasized & $(1,1,1,3)$   & $0.5 I_4$  & 0.0092 \\
\bottomrule
\end{tabular}
\end{table}

Overall, the results of Case Study~1 show how propagating preference uncertainty leads to a richer characterization of the decision space. The induced probability distribution over discrete extreme-point Pareto-optimal designs reveals which solutions are robustly favored under the assumed preference model, rather than treating all efficient designs as equally viable. The global sensitivity analysis further identifies which design variables drive variability in these selected solutions, indicating where targeted refinement may most effectively reduce decision uncertainty. Finally, the Fréchet variance summarizes the overall stability of the recommended design under a given preference distribution. Small values signal that preference uncertainty has limited impact on the final choice, whereas large values indicate substantial dispersion in the optimal decisions and may motivate additional preference elicitation.

\subsection{Case Study 2: Continuous optimal decision distribution}
\begin{figure}[H]
    \centering
    \includegraphics[width=1\textwidth]{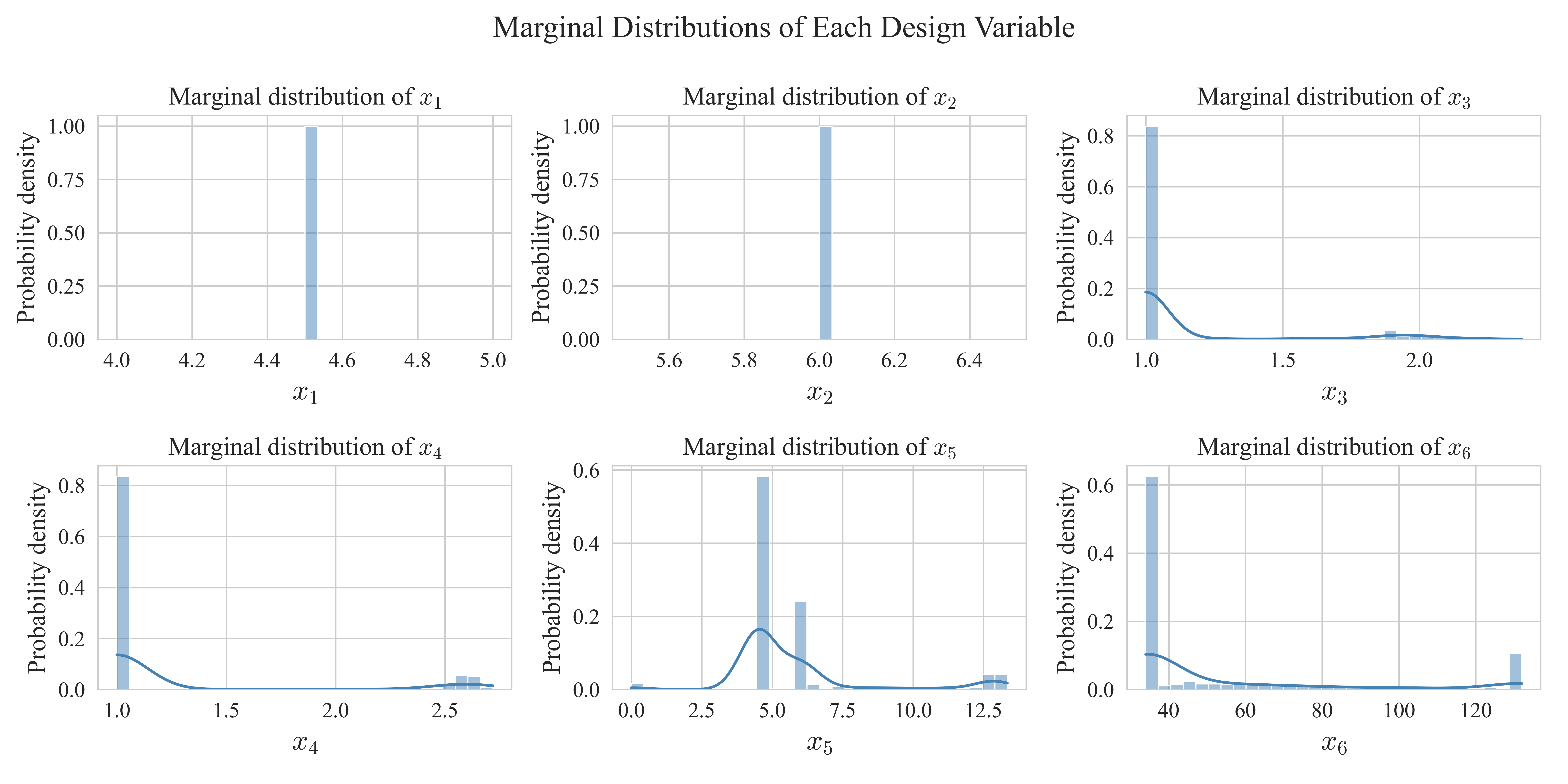}
    \caption{Marginal distributions of each decision variable for the continuous optimal decision distribution.}
    \label{fig:preference_problem3_norm_t_star_dist}
\end{figure}

In Case Study 2, the fourth objective function in \eqref{eq:f4} is defined as
\[
\tilde f_4 = \dfrac{\big[(x_6-2x_5-3.414\,x_3-0.586\,x_4)14\big]^2-1936}{3211264},
\]
which corresponds to the Pivot/Skid steering configuration. Unlike the LP setting in Case Study 1 where optima lie at extreme points, the convex quadratic formulation in Case Study 2 induces optimal solutions to vary continuously with the preference vector $\bm \beta$. Although the minimizer typically lies on the boundary of the feasible region due to the structure of the objectives, it is not restricted to a finite set of discrete points. Consequently, the induced optimal decision distribution is continuous, as reflected in the smooth marginal distributions of the decision variables shown in Figure~\ref{fig:preference_problem3_norm_t_star_dist}.

Since the scalarized problems in this case are convex quadratic programs, a fixed preference vector $\bm\beta$ typically yields a unique optimizer. Non-uniqueness, when it occurs, arises from degeneracy in the problem geometry rather than from the preference realization itself. As in Case Study~1, we retain the single representative solution returned by the solver for each sampled $\bm\beta$, so that the induced decision distribution, and the resulting sensitivity analysis and Fréchet variance, reflect variability driven by preference uncertainty rather than degeneracy artifacts. 

\begin{figure}[H]
    \centering
    \includegraphics[width=1\textwidth]{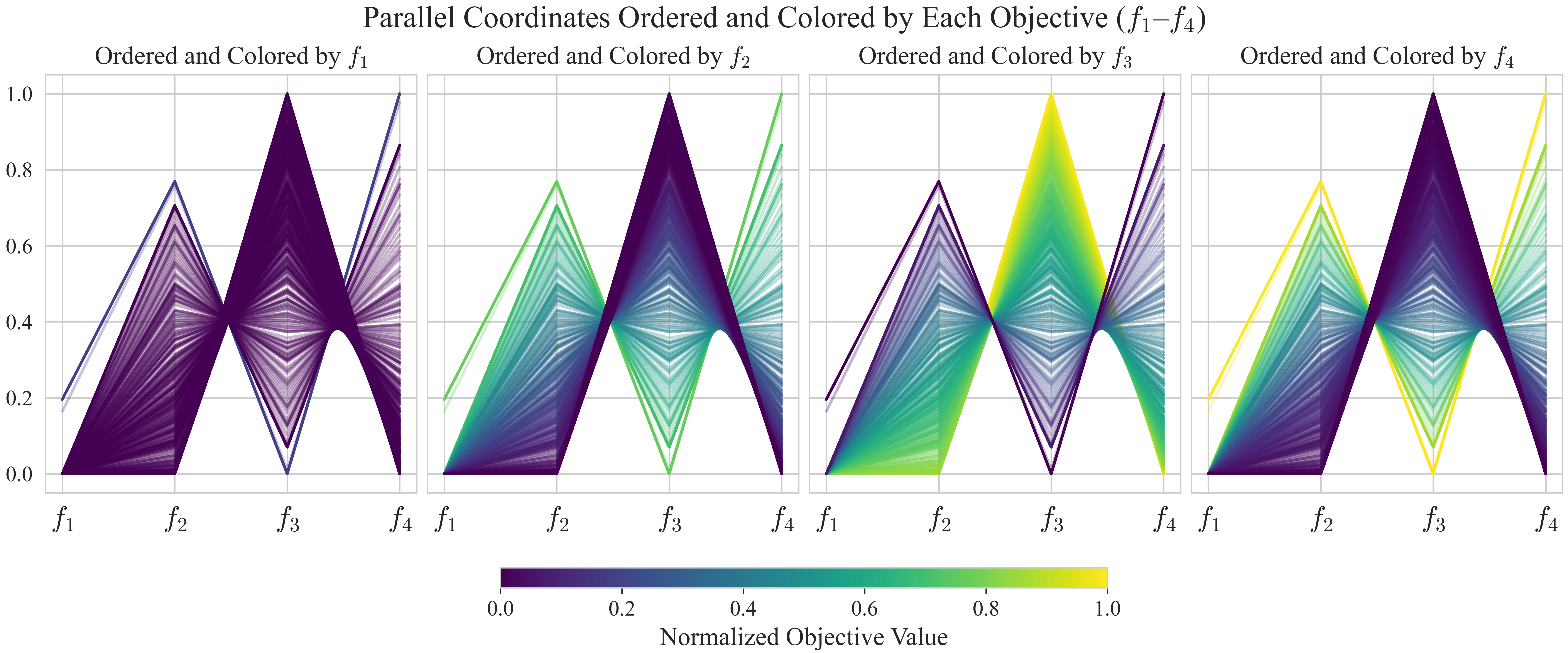}
    \caption{Parallel coordinates plot of normalized objective values for Pareto-optimal solutions in Case Study 2 (continuous distribution).}
    \label{fig:preference_problem3_norm_parallel_coordinates}
\end{figure}

As shown in Figure~\ref{fig:preference_problem3_norm_parallel_coordinates}, the Pareto-optimal solutions for the quadratic case display smooth and continuous trade-offs among the objectives. The variation in $f_1$ remains relatively limited, while $f_2$ and $f_4$ exhibit opposing trends with respect to $f_3$. In contrast with the discrete setting of Case Study 1, the trajectories here form continuous bands across the parallel coordinates, reflecting the underlying continuous distribution of optimal decisions. Furthermore, $f_2$ and $f_4$ share a broadly similar trend, suggesting consistent patterns of trade-off behavior.

\begin{figure}[H]
    \centering
    \includegraphics[width=1\textwidth]{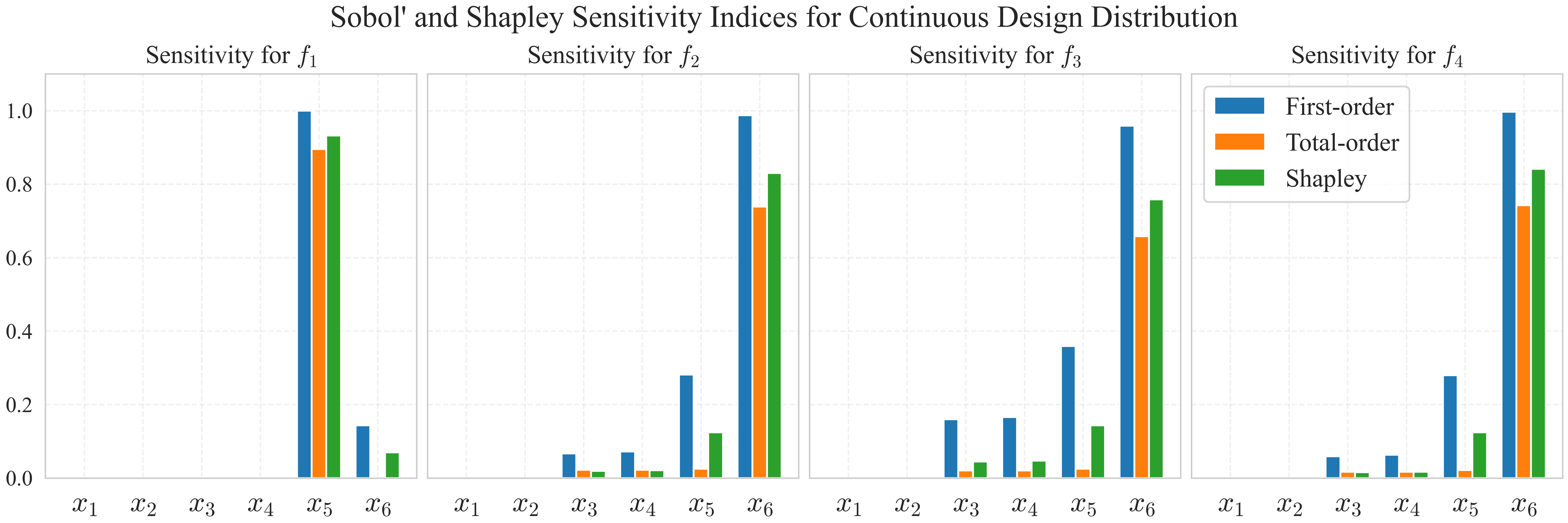}
    \caption{Sobol' and Shapley sensitivity indices for continuous optimal decision solutions across the four objectives.}
    \label{fig:preference_problem3_sobol_shapley}
\end{figure}

\begin{figure}[H]
    \centering
    \includegraphics[width=1\textwidth]{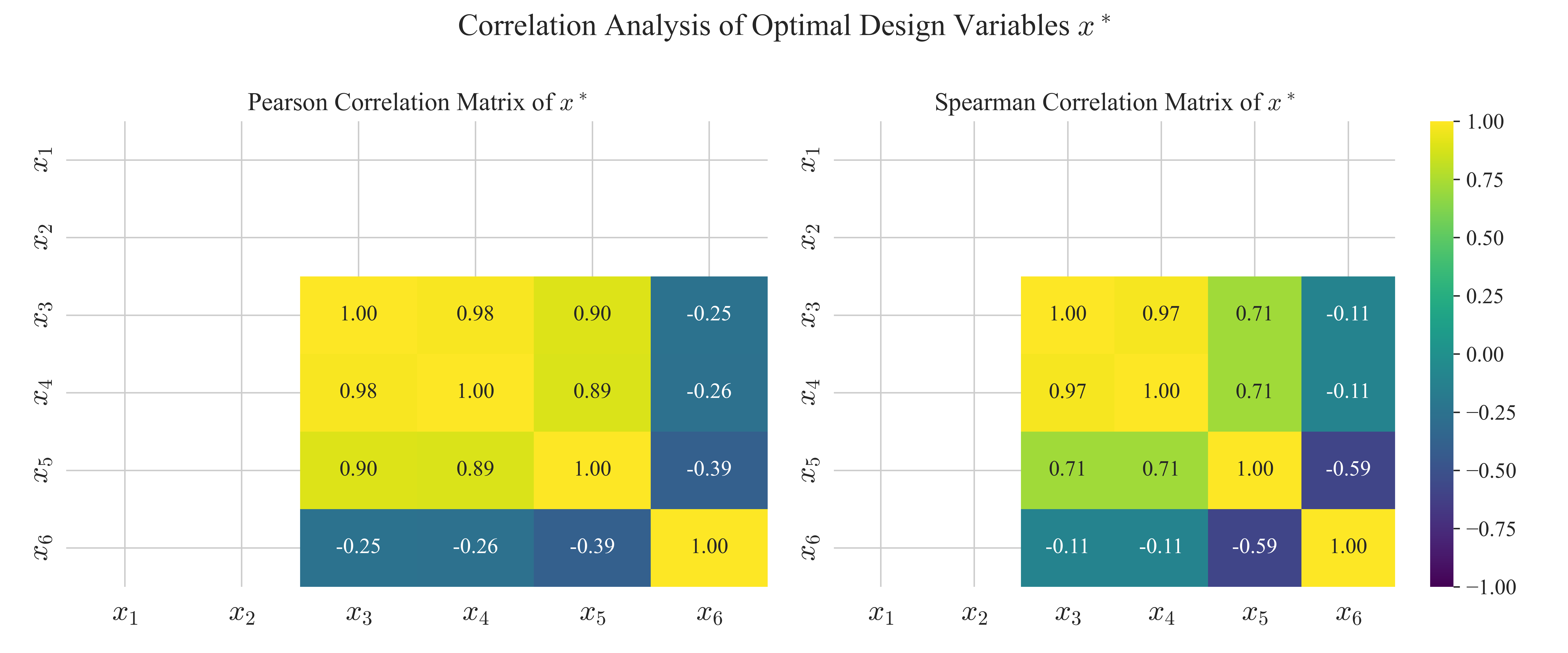}
    \caption{Pearson and Spearman correlation matrices of the continuous design distribution.}
    \label{fig:preference_problem3_norm_t_star_dependence}
\end{figure}

Figure~\ref{fig:preference_problem3_sobol_shapley} shows the Sobol' indices and Shapley values for the continuous optimal decision distribution. The results indicate that $x_5$ and $x_6$ exert the strongest influence across all objectives. For $f_1$, $x_5$ is dominant with only a minor contribution from $x_6$. For $f_2$–$f_4$, the effect of $x_6$ becomes more pronounced, while $x_3$ and $x_4$ contribute directly to $f_3$; since $f_4$ is a quadratic transformation of $f_3$, their influence naturally propagates to $f_4$. As in Case Study 1, we observe instances where first-order Sobol' indices exceed their total-order counterparts, reflecting the presence of dependencies among the design variables. These dependencies are further illustrated in Figure~\ref{fig:preference_problem3_norm_t_star_dependence}, which shows strong correlations among $x_3$, $x_4$, and $x_5$, as well as a negative association between $x_5$ and $x_6$. In such higher-dimensional settings with dependent inputs, Shapley indices are not constrained to lie between Sobol' first- and total-order indices. Rather, they redistribute joint contributions among correlated variables, yielding a more consistent attribution of variable importance under dependence.

\begin{figure}[H]
    \centering
    \includegraphics[width=1\textwidth]{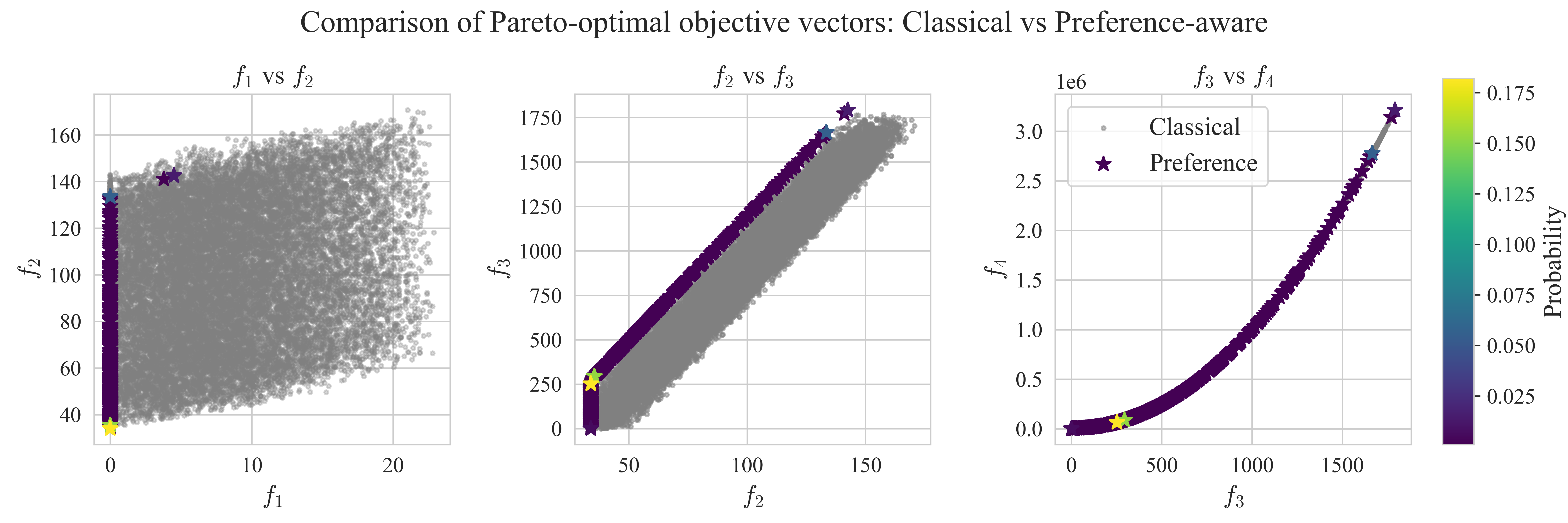}
    \caption{Classical Pareto-optimal objective vectors (grey) vs. preference-aware solutions (colored by probability mass) for continuous optimal decision.}
    \label{fig:preference_problem3_comparison_moo_preference}
\end{figure}

In contrast to classical approaches that approximate a Pareto front via LHS, our method induces a probability distribution over optimal decisions under uncertain preferences. While classical multi-objective optimization identifies a set of Pareto-efficient solutions, it provides no information about how often different regions of the front are selected under preference variability. Figure~\ref{fig:preference_problem3_comparison_moo_preference} visualizes this distinction. The grey points represent the Pareto-optimal objective vectors obtained from the classical baseline, whereas the colored points correspond to preference-aware optimal solutions, with color indicating empirical selection frequency. The results show that certain regions of the Pareto front are selected repeatedly across preference realizations, while other Pareto-efficient regions are realized only under specific preference configurations. This distributional view allows decision makers to identify regions of the Pareto front that are robustly favored under preference uncertainty and to distinguish them from regions that are efficient but highly sensitive to preference variation.

While the continuous distribution in Figure~\ref{fig:preference_problem3_norm_t_star_dist} illustrates how optimal designs vary across the feasible region under preference uncertainty, it is often helpful to summarize this dispersion with a single numerical measure. The Fréchet variance provides such an aggregate indicator by quantifying how widely the optimal solutions vary across repeated draws of the preference vector $(\bm \mu, \bm \Sigma)$. Since the solutions move smoothly across the Pareto-optimal region of the design space, different preference distributions can induce substantially different degrees of spread. Small Fréchet variance values indicate that most sampled preferences concentrate the optimal designs within a narrow portion of the feasible region, whereas larger values signify that the induced solutions occupy a broader subset of the front. Table~\ref{tab:frechet_variance_continuous} reports these values for several representative preferences.

\begin{table}[H]
\centering
\caption{Fréchet variance of the continuous optimal decision distribution under different preference settings.}
\label{tab:frechet_variance_continuous}
\begin{tabular}{lccS[table-format=1.3]}
\toprule
Scenario & $\bm \mu$ & $\bm \Sigma$ & {Fréchet variance} \\
\midrule
Baseline          & $(1,1,1,1)$   & $0.5 I_4$  & 0.2284 \\
Low uncertainty   & $(1,1,1,1)$   & $0.1 I_4$  & 0.0265 \\
High uncertainty  & $(1,1,1,1)$   & $1.0 I_4$  & 0.3136 \\
Objective 3 emphasized & $(1,1,3,1)$   & $0.5 I_4$  & 0.1073 \\
Objective 4 emphasized & $(1,1,1,3)$   & $0.5 I_4$  & 0.1886 \\
\bottomrule
\end{tabular}
\end{table}

Overall, the results of Case Study~2 illustrate how propagating preference uncertainty yields a detailed, probabilistic characterization of the continuous design space. The induced distribution reveals which regions of the continuous Pareto front are most likely to be selected under the assumed preference model, as opposed to treating all efficient solutions as equally viable. The sensitivity analysis further identifies which design variables jointly drive the continuous variation in these optimal decisions, providing guidance on where targeted adjustments may most effectively reduce decision uncertainty. The Fréchet variance summarizes this stability in a single scalar. Small values indicate that most preference realizations lead to tightly clustered designs, whereas large values reflect a wide dispersion of the optimal design distribution. When the dispersion is large, the breadth of the resulting design set suggests that the current preference specification is not sufficiently informative, and that collecting additional preference information may help narrow the range of credible recommendations.

\section{Conclusion}\label{s:conclusion}
This study introduces a probabilistic framework for multi-objective engineering design in which preferences are modeled as random variables. Under this setting, optimal solutions become random outcomes, inducing a probability distribution over the design space. This perspective goes beyond classical Pareto analysis by not only identifying non-dominated solutions, but also quantifying how likely each solution is to be selected under uncertain preferences. We demonstrate the framework using two vehicle design problems. In the Ackermann steering case, the problem reduces to an LP and yields a discrete distribution over a finite set of extreme-point designs. In contrast, the pivot/skid steering case produces a continuous distribution of optimal solutions due to its nonlinear structure. These results show that preference uncertainty can lead to fundamentally different forms of decision uncertainty depending on the problem structure. Furthermore, global sensitivity analysis reveals how variability in each design variable contributes to the induced decision distribution, capturing both independent and correlated effects.

The proposed framework offers decision makers richer insight than classical multi-objective optimization by explicitly linking preference uncertainty to decision uncertainty. Instead of providing only a Pareto front, it quantifies how likely each design is to be realized, which supports risk-aware and interpretable decision-making. The induced distribution also provides a natural indicator of when additional preference elicitation may be valuable, since highly dispersed decision distributions suggest that the current preference model is not sufficiently informative to yield stable recommendations. A natural extension of this work is to leverage the induced optimal-solution distribution for preference elicitation and interactive design. In particular, the observed distribution of preferred solutions can serve as informative feedback to update or infer decision-makers' preferences, forming a bridge between optimization and learning. Future research may investigate how the induced optimal-solution distribution can support sequential or optimal-stopping strategies for preference elicitation, in which additional queries are issued only when the current level of decision uncertainty remains too high to make reliable design recommendations.

% \section*{Supplementary materials}

\if0\blind{
\section*{Acknowledgements}
This work was supported by Clemson University's Virtual Prototyping of Autonomy Enabled Ground Systems (VIPR-GS), under Cooperative Agreement W56HZV-21-2-0001 with the US Army DEVCOM Ground Vehicle Systems Center (GVSC).
DISTRIBUTION STATEMENT A. Approved for public release; distribution is unlimited. OPSEC10263
} 
\fi

\section*{Data Availability Statement}
Our code and data are available in a GitHub repository: \url{https://github.com/CRLiu0818/Estimating_Decision_Uncertainty_from_Preference_Uncertainty.git}.

\bibliographystyle{chicago}
\spacingset{1}
\bibliography{IISE-Trans}

\end{document}